\documentstyle[aps,prb,twocolumn,epsf,rotate]{revtex}

\begin{document}

\twocolumn[\hsize\textwidth\columnwidth\hsize\csname
@twocolumnfalse\endcsname

\title{ Pseudospin Anisotropy Classification of  Quantum Hall Ferromagnets}

\draft

\author{T. Jungwirth$^{1,2}$ and A.H. MacDonald$^{1}$}
\address{$^{1}$Department of Physics,
Indiana University, Bloomington, Indiana 47405}
\address{$^{2}$Institute of Physics ASCR,
Cukrovarnick\'a 10, 162 00 Praha 6, Czech Republic}
\date{\today}
\maketitle

\begin{abstract}

Broken symmetry ground states with uniform electron density
are common in quantum Hall systems
when two Landau levels simultaneously approach the chemical potential
at integer filling factor $\nu$.  The close analogy between these
two-dimensional electron system states and conventional itinerant electron
ferromagnets can be emphasized by using a pseudospin label to distinguish the
two Landau levels.  As in conventional ferromagnets, the evolution
of the system's state as external field parameters are varied
is expected to be sensitive to the dependence of ground state energy on
pseudospin orientation.  We discuss the predictions of Hartree-Fock
theory for the dependence of the sign and magnitude of the
pseudospin anisotropy energy on the nature of the crossing
Landau levels.  We build up a classification scheme for
quantum Hall ferromagnets that applies for single layer and
bilayer systems with two aligned Landau levels distinguished
by any combination of real-spin, orbit-radius, or growth direction
degree-of-freedom quantum numbers.  The possibility of in-situ tuning
between easy-axis and easy-plane quantum Hall ferromagnets is discussed for
biased bilayer systems with total filling factors $\nu=3$ or $\nu=4$.
Detailed predictions are made for the bias dependence of
pseudospin reversal properties in $\nu=3$ bilayer systems.

\end{abstract}

\pacs{73.40.Hm, 75.10.Lp, 75.30.Gw}


\vskip2pc]
\section{Introduction}
Studies of magnetic phenomena in semiconductors have opened fruitful
new ways to explore the subtleties of quantum magnetism.
In quantum Hall samples, the tunability of semiconductor electronic
systems and the quantization of single-particle energies into macroscopically
degenerate Landau levels (LLs) combine to open up a rich and varied phenomenology.
The effective zero width of electronic energy
bands enhances the role of inter-particle
interactions\cite{ahmintro} and can frequently lead to the formation of ordered
many-particle ground states, including ferromagnetic ones.

Most studies of quantum Hall
ferromagnets\cite{dassarmabook} (QHFs ) have focused on
spontaneous spin-alignment in a single-layer two-dimensional (2D)
electron system at LL filling factor
$\nu=1$. In this case it turns out that
electron-electron interactions favor
fully aligned electron spins even in the limit of vanishingly small Zeeman
coupling\cite{dassarmabook,qhferro}
and the ferromagnetic ground state of the system is described exactly by
Hartree-Fock (HF) theory.  Because
of the near spin-independence of the Coulomb interaction, the
$\nu=1$ single layer QHF is a Heisenberg-like isotropic two-dimensional
ferromagnet.  One of the unique properties of this simple itinerant
electron ferromagnet is that its instantons\cite{qhferro}
(Skyrmions) carry charge and can be observed\cite{skyrmexps} in the
ground states at filling factors slightly deviating
from 1.

The notion of the QHF  can be generalized however.
It turns out that, at least according to HF theory,
broken symmetry ground states occur at integer filling factors
in quantum Hall systems any time two or more valence LLs are degenerate and
the number of electrons is sufficient to fill only some of the LLs.
In effect, electrons in the ordered state occupy spontaneously generated
LLs that are linear combinations of the single-particle levels chosen to
minimize the electron interaction energy.

The simplest example of a pseudospin QHF obtains at $\nu=1$ in balanced
bilayer 2D systems where the single-particle LLs in the
two-layers are degenerate.
In the ordered ground state\cite{dassarmabook,ahm90,dltheory}
the electrons occupy a LL which is a linear combination of the
isolated layer levels, forming a state with spontaneous inter-layer phase coherence.
Recently, Josephson-like behavior seen\cite{eisensteinlatest} in
2D-to-2D tunneling spectroscopy studies of bilayer systems has
provided a  direct manifestation of collective behavior
generated by this broken symmetry.
In $\nu=1$ bilayer systems, the broken symmetry state
minimizes the Hartree energy cost by distributing charge equally between
layers and gives up part of the intra-layer exchange energy while gaining
more in inter-layer exchange energy. Unlike the single layer $\nu=1$ QHF,
the ordered HF ground state is not exact in this case, and the order
predicted by HF theory can be destroyed.
With decreasing inter-layer exchange energy quantum fluctuations around
the mean-field ordered state become more important and for layer
separations larger than approximately two magnetic lengths fluctuations
destroy the spontaneous coherence. The corresponding order-disorder quantum
phase transition has been observed experimentally.\cite{murphy}
In $\nu=1$ bilayer QHFs, it is the layer degree of freedom
which is represented as a {\em pseudospin}-1/2.
With this mapping the phase coherent state is equivalent to
a spin-1/2 easy-plane ferromagnet. Finite temperature Kosterlitz-Thouless
phase transition\cite{kt},
continuous quantum phase transition induced by in-plane magnetic
field\cite{murphy}, and macroscopic collective transport effects\cite{eisensteinlatest}
are are among the remarkable phenomena which have been studied on bilayer
QHFs. 

Recent experiments in single layer\cite{mansour} and bilayer\cite{vittorio}
2D systems at even-integer filling factors
have further enlarged the field of quantum Hall
ferromagnetism. It has been shown that easy-axis ferromagnetic
ground states can occur\cite{mansour} at higher filling factors when LLs with
different orbit radius quantum numbers are brought close to
alignment.\cite{giuliani,mansour}
In HF language, easy-axis anisotropy means that many-body states with
either of the two aligned LLs completely filled and the
other empty is energetically more favorable than the  coherent superposition state.
The easy-axis pseudospin anisotropy occurs in this case because intra LL exchange
is stronger than exchange between particles from LLs with different orbit radius quantum
numbers.  Transport measurements\cite{vittorio,woowon}
have demonstrated that easy-axis QHFs exhibit hysteresis with a complicated
phenomenology, presumably associated with an interplay between disorder and
domain-morphology similar to that in conventional thin film magnets.

In this paper we classify QHFs according to their
pseudospin anisotropy energies as either isotropic, easy-axis, or easy-plane
systems.  We report on a HF based analysis which predicts
how the class of broken symmetry ground state depends on the nature of
the crossing LLs.  In some cases, competing effects allow the
anisotropy energy to be tuned continuously generating a zero-temperature
quantum phase transition between different classes of states.
We consider only cases where no more than two LLs are nearly
degenerate and the number of electrons is sufficient to occupy one of
them.  We will always assume that lower energy LLs, if present, are completely full and
higher energy LLs are completely empty; coupling to these remote
LLs can usually be treated perturbatively if necessary, although
we do not do so explicitly here.  An important example of an instance in
which more than two LLs are close to degeneracy pertains 
in double-layer systems with weak tunneling and weak Zeeman
coupling;\cite{fourcomptheory,fourcompexp} 
we do not treat this or other more complex cases with many
degenerate LLs in this paper.
In Section~II we precisely define the pseudospin language we use
in which one of the LLs is referred to as the pseudospin-up state and  the
other LL as the pseudospin-down state.  Since we assume that the magnetic field
is perpendicular to the 2D electron layer, growth direction and in-plane degrees
of freedom decouple.  The pseudospin quantum number then subsumes
real-spin, orbit radius, and growth direction (subband)
degrees of freedom. To make the discussion more transparent we concentrate on
a system consisting of two nearby infinitely narrow 2D layers,
the simplest model which has a non-trivial growth direction degree of freedom.
Comments are made throughout the text about realistic samples
with more complicated geometries. In Section~III
we derive a general expression for the HF ground state energy in the pseudospin
ferromagnetic state.
Section~IV summarizes the rather cumbersome evaluation of Coulomb interaction
matrix elements.  Readers not interested
in technical details of the calculation are encouraged to skip to Section~V
where we present our conclusions concerning pseudospin magnetic anisotropy
of single layer and bilayer QHFs.  This section includes phase diagrams
which show the regimes of physically tunable parameters with
easy-axis and easy-plane anisotropies.
Symmetry breaking fields and the dynamics of pseudospin
reversal are discussed in Section~VI.  Finally, we conclude in Section~VII  with a
brief summary of the main results of our paper.
\section{Pseudospin representation}
The pseudospin language was introduced\cite{ahm90}
to the description of broken symmetry states in the quantum Hall regime,
in order to draw on the analogy between double quantum well systems
at $\nu=1$ and 2D ferromagnets.  In this work the pseudospin degree of freedom
represented the layer index of a bilayer system.
Here we allow the pseudospin index to have a more general meaning.
To establish terminology, it is useful to recall the
single-particle spectrum of a bilayer 2D system subject to
a perpendicular magnetic field.
Quantum well subbands of individual layers can be mixed by
interlayer tunneling and shifted by the application of
a bias potential.  We limit our attention to the usual case where only
the lowest electric subband of either quantum well is occupied and
for explicit calculations use a zero-width quantum well model.
Allowing for external bias and for tunneling between the wells
(see Fig.~\ref{dqw}), the bilayer subband wavefunctions of the
zero-width model are
\begin{equation}
\lambda_{\pm 1}(z)=\frac{1}{\sqrt{2}}\left[ (1\mp r_{\Delta})^{1/2}\delta (z)
\pm(1\pm r_{\Delta})^{1/2}\delta (z-d)\right ] \;,
\label{zwf}
\end{equation}
where $r_{\Delta}=\Delta_V/(\Delta_V^2+\Delta_t^2)^{1/2}$,  $\Delta_V$
is the bias potential, $\Delta_t$ the tunneling gap at zero bias, and $d$ is
the layer separation.
To account for  specific experimental samples,
finite width effects can be incorporated by replacing the  
wavefunctions (\ref{zwf}) with electric subband wavefunctions 
calculated using the self-consistent
local-spin-density-approximation (LSDA) model.\cite{bastard}

In the Landau gauge the wavefunctions in the 2D plane take a form
$\phi_{n,s,k}(x)\exp(iky)/\sqrt{L_y}$, where
\begin{eqnarray}
\phi_{n,s,k}(x)&=&
\big[\pi\ell^22^{2n}(n!)^2\big]^
{-1/4}H_n\left(\frac{x-\ell^2k}{\ell}\right)\nonumber \\
&\times&\exp\left[-\frac{(x-\ell^2k)^2}{2\ell^2}\right]\; ,
\label{xywf}
\end{eqnarray}
$k$ is the wavevector label which distinguishes states within a
LL, $n=0,1,...$ is the orbit radius quantum number, $\ell$ is the
magnetic length,
and we have also explicitly included the real-spin,
$s=\pm1/2$, degree of freedom. The single-particle energy spectrum
consists of discrete LLs
\begin{equation}
E_{\xi,n,s}=-\frac{\xi}{2}
(\Delta_V^2+\Delta_t^2)^{1/2}+\hbar\omega_c(n+\frac12)
-s|g|\mu_BB\;,
\label{ll}
\end{equation}
where $\omega_c$ is the cyclotron frequency  and the last term is the
real-spin Zeeman coupling. Each LL has a
macroscopic degeneracy with the
number of orbital states per level
$N_{\phi}=AB/\Phi_0$, where $A$ is the system area, $B$ is
the field strength, and $\Phi_0$ is the magnetic flux quantum.

The many-body broken symmetry states we study in the following sections occur
when two LLs are brought close to alignment while remaining sufficiently
separated from other LLs.  In our calculations, each LL can have one
of two possible subband indices, one of two possible spin indices, and
any value for the 2D cyclotron orbit kinetic energy index.
The two crossing LLs can differ in any or all of these labels.
We label one of the two levels as the pseudospin-up
($\sigma=\uparrow$) state and the other level as the pseudospin down
($\sigma=\downarrow$) state.  We truncate the single-particle Hilbert space
by ignoring higher LLs and introducing effective
one-body fields that account for the  effect
of electrons in lower LLs on the two pseudospin states.
Within this model the set of single-particle states reduces
to following wavefunctions
\begin{equation}
\psi_{\sigma,k}(\vec{r})=\lambda_{\xi(\sigma)}(z)\;\phi_{n(\sigma),s(\sigma),
k}(x)\;\frac{\exp(iky)}{\sqrt{L_y}}\; .
\label{pswf}
\end{equation}
A particle with the pseudospin oriented along a general
unit vector $\hat{m}=(\sin{\theta}
\cos{\varphi},\sin{\theta}\sin{\varphi},\cos{\theta})$
is described by
\begin{equation}
\psi_{\hat{m},k}(\vec{r})=\cos\big(\frac{\theta}{2}\big)\psi_{\uparrow,k}(\vec{r})
+\sin\big(\frac{\theta}{2}\big)e^{i\varphi}\psi_{\downarrow,k}(\vec{r})\; .
\label{generalps}
\end{equation}
\section{Many-body Hamiltonian and HF total energy}
In the HF approximation, the
QHF has a single Slater determinant state
with the same pseudospin orientation for every orbital $k$.
In this section we derive general expressions for the dependence
of the many-electron state energy on pseudospin orientation.
It is convenient to
express the many-body Hamiltonian using Pauli spin-matrices $\tau_x$,
$\tau_y$, and $\tau_z$ and the $2\times 2$ identity matrix which we label
$\tau_{\protect\bf 1}$.
In this representation the Hamiltonian reads
\begin{eqnarray}
H&=&-\sum_{i={\protect\bf 1},x,y,z}\sum_{k=1}^{N_{\phi}}\sum_{\alpha,\alpha^{\prime}
=1}^{2}b_i\tau_i^{\alpha^{\prime},\alpha}c^{\dagger}_{\sigma(\alpha^{\prime}),k}
c_{\sigma(\alpha),k}\nonumber\\
&+&\frac12 \sum_{i,j={\protect\bf 1},x,y,z}\sum_{k_1,k^{\prime}_1,
\atop k_2,
k^{\prime}_2=1}^{N_{\phi}}\sum_{\alpha_1,\alpha^{\prime}_1,\atop
\alpha_2,\alpha^{\prime}_2=1}^{2}W_{i,j}^{k^{\prime}_1,k^{\prime}_2,
k_1,k_2}\tau_i^{\alpha^{\prime}_1,\alpha_1}\tau_j^{\alpha^{\prime}_2,\alpha_2}
\nonumber \\
&\times&
c^{\dagger}_{\sigma(\alpha^{\prime}_1),k^{\prime}_1}
c^{\dagger}_{\sigma(\alpha^{\prime}_2),k^{\prime}_2}
c_{\sigma(\alpha_2),k_2}
c_{\sigma(\alpha_1),k_1}\; ,
\label{hamiltonian}
\end{eqnarray}
where $\sigma(1)=\uparrow$ and $\sigma(2)=\downarrow$.
The one-body terms $b_i$ include, in general, the external bias potential,
tunneling, cyclotron and Zeeman energies and also the mean-fields from
interactions with electrons in the  frozen LLs lying below
the $\sigma=\uparrow$ and $\downarrow$ levels.
We will give an explicit expression for $b_i$ in  section~VI.
Here and in the following
two sections, we concentrate on the
two-body terms in the Hamiltonian~(\ref{hamiltonian}).

The potentials $W_{i,j}$ represent different combinations of Coulomb
interaction matrix elements, $V_{\sigma_1^{\prime},
\sigma_2^{\prime},\sigma_1,\sigma_2}$,
of the single-particle pseudospin states
\begin{eqnarray}
V_{\sigma_1^{\prime},
\sigma_2^{\prime},\sigma_1,\sigma_2}^{
k^{\prime}_1,k^{\prime}_2,
k_1,k_2}&=&\int d^3\vec{r}_1\int d^3\vec{r}_2
\psi^*_{\sigma_1^{\prime},k^{\prime}_1}(\vec{r}_1)
\psi^*_{\sigma_2^{\prime},k^{\prime}_2}(\vec{r}_2)\nonumber \\
&\times& \frac{e^2}{\epsilon|\vec{r}_1-\vec{r}_2|}
\psi_{\sigma_1,k_1}(\vec{r}_1)
\psi_{\sigma_2,k_2}(\vec{r}_2)
\label{vssss}
\end{eqnarray}
General expressions for the pseudospin dependent interactions $W_{i,j}$
are given in Table~\ref{wij} in terms of the following
matrix element combinations:
\begin{eqnarray}
B_1^{\pm}&=&\frac14(V_{\uparrow,\uparrow,\uparrow,\uparrow}\pm
V_{\downarrow,\downarrow,\downarrow,\downarrow})\; ,
B_2^{\pm}=\frac14(V_{\uparrow,\downarrow,\uparrow,\downarrow}\pm
V_{\downarrow,\uparrow,\downarrow,\uparrow})\nonumber \\
B_3^{\pm}&=&\frac14(V_{\uparrow,\downarrow,\downarrow,\uparrow}\pm
V_{\downarrow,\uparrow,\uparrow,\downarrow})\; ,
B_4^{\pm}=\frac14(V_{\uparrow,\uparrow,\downarrow,\downarrow}\pm
V_{\downarrow,\downarrow,\uparrow,\uparrow})\nonumber \\
B_5^{\pm}&=&\frac14(V_{\uparrow,\uparrow,\uparrow,\downarrow}\pm
V_{\uparrow,\downarrow,\uparrow,\uparrow})\; ,
B_6^{\pm}=\frac14(V_{\downarrow,\uparrow,\downarrow,\downarrow}\pm
V_{\downarrow,\downarrow,\downarrow,\uparrow})\nonumber \\
B_7^{\pm}&=&\frac14(V_{\uparrow,\uparrow,\downarrow,\uparrow}\pm
V_{\downarrow,\uparrow,\uparrow,\uparrow})\; ,
B_8^{\pm}=\frac14(V_{\uparrow,\downarrow,\downarrow,\downarrow}\pm
V_{\downarrow,\downarrow,\uparrow,\downarrow})\nonumber \\
\label{b}
\end{eqnarray}
In (\ref{b}) we have omitted orbital guiding center indices for simplicity.

The many-electron state with pseudospin orientation
$\hat m$ is $|\Psi[\hat m] \rangle =
\prod_{k=1}^{N_{\phi}} c^{\dagger}_{\hat m,k} |0\rangle$,
where $c^{\dagger}_{\hat m,k}$ creates the single-particle
state whose wavefunction is given in Eq.~(\ref{generalps}).
We find that
\begin{eqnarray}
e_{HF}(\hat m) &\equiv&\frac{\langle \Psi[\hat m] | H | \Psi[\hat m] \rangle }
{N_{\phi}}\nonumber \\
&=&
-\sum_{i=x,y,z}\left(b_i-
\frac12U_{{\protect\bf 1},i}-\frac12U_{i,{\protect\bf 1}}\right)m_i
\nonumber \\
&+&\frac12\sum_{i,j=x,y,z}U_{i,j}m_im_j\; ,
\label{ehf}
\end{eqnarray}
where
\begin{equation}
U_{i,j}=\frac{1}{N_{\phi}}\sum_{k_1,k_2=1}^{N_{\phi}}\left(
W_{i,j}^{k_1,k_2,k_1,k_2}-W_{i,j}^{k_2,k_1,k_1,k_2}\right)
\label{uij}
\end{equation}
has direct and exchange contributions.  Eq.~(\ref{ehf}) is the most
general form for the HF energy of a pseudospin-1/2 QHF. The magnetic anisotropy
of the particular QHF system is governed by the terms in~(\ref{ehf})
that are quadratic in the pseudospin magnetization $m_i$. The values of
the coefficients $U_{i,j}$ depend on the nature of the crossing LLs and
and can result in isotropic, easy-plane, or easy-axis quantum Hall ferromagnetism.
\section{Pseudospin matrix elements of the Coulomb interaction}
This section contains  the derivation of explicit expressions for
the anisotropy energy coefficients $U_{i,j}$, assuming the zero-width
quantum well model wavefunctions (\ref{zwf}).
Using the Fourier representation of the Coulomb interaction
we write the pseudospin matrix elements as
\begin{eqnarray}
&&V_{\sigma_1^{\prime},
\sigma_2^{\prime},\sigma_1,\sigma_2}^{
k^{\prime}_1,k^{\prime}_2,
k_1,k_2}=
\frac1A\sum_{\vec{q}}\delta_{q_y,k_1^{\prime}-k_1}
\delta_{-q_y,k_2^{\prime}-k_2}\nonumber \\
&\times&
e^{iq_x(k_1^{\prime}+k_1)/2}
e^{-iq_x(k_2^{\prime}+k_2)/2}
v_{\sigma_1^{\prime},
\sigma_2^{\prime},\sigma_1,\sigma_2}(\vec q)
\label{fourier}
\end{eqnarray}
and hence
\begin{eqnarray}
&&\frac{1}{N_{\phi}}\sum_{k_1,k_2=1}^{N_{\phi}}\big(
V_{\sigma_1^{\prime},
\sigma_2^{\prime},\sigma_1,\sigma_2}^{k_1,k_2,k_1,k_2}
-V_{\sigma_1^{\prime},
\sigma_2^{\prime},\sigma_1,\sigma_2}^{k_2,k_1,k_1,k_2}\big)
=\nonumber \\
&&\frac{N_{\phi}}{A}v_{\sigma_1^{\prime},
\sigma_2^{\prime},\sigma_1,\sigma_2}(0)
-
\frac1A\sum_{\vec q}
v_{\sigma_1^{\prime},
\sigma_2^{\prime},\sigma_1,\sigma_2}(\vec q)
\label{sumq}
\end{eqnarray}
In the following we take $\ell$ as the unit of length
and  $e^2/\epsilon\ell$ as the unit of energy. Then
$N_{\phi}/A=1/2\pi$ and Eq.~(\ref{sumq}) can be rewritten in the
more transparent form
\begin{eqnarray}
&&\frac{1}{N_{\phi}}\sum_{k_1,k_2=1}^{N_{\phi}}\big(
V_{\sigma_1^{\prime},
\sigma_2^{\prime},\sigma_1,\sigma_2}^{k_1,k_2,k_1,k_2}
-V_{\sigma_1^{\prime},
\sigma_2^{\prime},\sigma_1,\sigma_2}^{k_2,k_1,k_1,k_2}\big)
=\nonumber \\
&&\int\frac{d^2\vec q}{(2\pi)^2}e^{-q^2/2}
\big[v_{\sigma_1^{\prime},
\sigma_2^{\prime},\sigma_1,\sigma_2}(0)-
v_{\sigma_1^{\prime},
\sigma_2^{\prime},\sigma_1,\sigma_2}(\vec q)\big]\;.
\label{intq}
\end{eqnarray}
The first factor in square brackets in this equation originates
from the Hartree contribution to the energy while the second
factor originates from the exchange contribution.

In these equations $v_{\sigma_1^{\prime},
\sigma_2^{\prime},\sigma_1,\sigma_2}(\vec q)$ is a pseudospin-dependent
effective 2D interaction which, because of
the separability of the in-plane and out-of-plane
degree-of-freedom terms in the single electron Schroedinger
equation, is the product of two factors: the subband factor
\begin{eqnarray}
&&v^{\Xi}_{\sigma_1^{\prime},
\sigma_2^{\prime},\sigma_1,\sigma_2}(\vec q)=
\int_{-\infty}^{\infty}dz_1\int_{-\infty}^{\infty}dz_2
e^{-q|z_1-z_2|}\nonumber \\
&\times&\lambda_{\xi(\sigma_1^{\prime})}(z_1)\,
\lambda_{\xi(\sigma_2^{\prime})}(z_2)\,
\lambda_{\xi(\sigma_1)}(z_1)\,
\lambda_{\xi(\sigma_2)}(z_2)
\label{vxi}
\end{eqnarray}
and the in-plane term
\begin{eqnarray}
&&v^{N}_{\sigma_1^{\prime},
\sigma_2^{\prime},\sigma_1,\sigma_2}(\vec q)=
e^{q^2/2}\nonumber \\
&\times&
\int_{-\infty}^{\infty}dx_1\,
\phi_{n(\sigma_1^{\prime}),s(\sigma_1^{\prime}),q_y/2}(x_1)\,
\phi_{n(\sigma_1),s(\sigma_1),-q_y/2}(x_1)\nonumber \\
&\times&
\int_{-\infty}^{\infty}dx_2\,
\phi_{n(\sigma_2^{\prime}),s(\sigma_2^{\prime}),-q_y/2}(x_2)\,
\phi_{n(\sigma_2),s(\sigma_2),q_y/2}(x_2)
\label{vn}
\end{eqnarray}

For a general sample geometry the subband factor (\ref{vxi}) has to be
calculated numerically using the self-consistent LSDA wavefunctions.
For our model bilayer system, however, we can obtain analytic expressions for
$v^{\Xi}_{\sigma_1^{\prime},
\sigma_2^{\prime},\sigma_1,\sigma_2}(\vec q)$.
In the  case of $\xi(\sigma)=\xi(-\sigma)$,
Eqs.~(\ref{zwf}) and (\ref{vxi}) give
\begin{equation}
v^{\Xi}_{\sigma_1^{\prime},\sigma_2^{\prime},\sigma_1,\sigma_2}=
\frac12\big[(1+r_{\Delta}^2)+
(1-r_{\Delta}^2)e^{-dq}\big]\;.
\label{vxisame}
\end{equation}
In the second case, i.e. $\xi(\sigma)=-\xi(-\sigma)$,
\begin{eqnarray}
&&v^{\Xi}_{\sigma,\sigma,\sigma,\sigma}=\frac12\big[(1+r_{\Delta}^2)+
(1-r_{\Delta}^2)e^{-dq}\big]\;, \nonumber \\
&&v^{\Xi}_{\sigma,-\sigma,\sigma,-\sigma}=\frac12\big[(1-r_{\Delta}^2)+
(1+r_{\Delta}^2)e^{-dq}\big]\;, \nonumber \\
&&v^{\Xi}_{\sigma,-\sigma,-\sigma,\sigma}=\frac12(1-r_{\Delta}^2)
(1-e^{-dq})\;, \nonumber \\
&& \nonumber \\
&& {\rm and}\nonumber \\
&& \nonumber \\
&&v^{\Xi}_{\sigma_1^{\prime},\sigma_2^{\prime},\sigma_1,\sigma_2}=
\eta\frac{r_{\Delta}}{2}
(1-r_{\Delta}^2)^{1/2}(1-e^{-dq})\nonumber \\
&& {\rm if}\;\;
\eta\equiv\frac12\sum_{i=1}^2\big[\xi(\sigma_i^{\prime})+\xi(\sigma_i)
\big]=\pm1\; .
\label{vxidiff}
\end{eqnarray}

For the in-plane factor in the effective interaction
it is necessary to distinguish several cases. If the pseudospin-up and pseudospin-down
levels have the same
real-spin index and same orbit-radius quantum number, i.e. $s(\sigma)=s(-\sigma)$
and $n(\sigma)=n(-\sigma)\equiv n$,
the effective interaction is independent of the pseudospin indices and
we obtain from (\ref{xywf}) and (\ref{vn})
\begin{equation}
v^{N}_{\sigma_1^{\prime},\sigma_2^{\prime},\sigma_1,\sigma_2}=
\frac{2\pi}{q}\big[L_n(q^2/2)\big]^2\; ,
\label{vnsamesame}
\end{equation}
where $L_n(x)$ is the Laguerre polynomial.
For identical spins but different orbit-radius quantum numbers, i.e.
$s(\sigma)=s(-\sigma)$ and $n(\sigma)\neq n(-\sigma)$, we define
$n_{<}\equiv\min\big[n(\sigma),n(-\sigma)\big]$ and
$n_{>}\equiv\max\big[n(\sigma),n(-\sigma)\big]$.  Then this factor in the effective
interaction is
\begin{eqnarray}
&& v^N_{\sigma,\sigma,\sigma,\sigma}=\frac{2\pi}{q}\big[L_{n(\sigma)}
(q^2/2)\big]^2\; , \nonumber \\
&& v^N_{\sigma,-\sigma,\sigma,-\sigma}=\frac{2\pi}{q}L_{n(\sigma)}
(q^2/2)L_{n(-\sigma)}(q^2/2)\; , \nonumber \\
&& v^N_{\sigma,-\sigma,-\sigma,\sigma}=\frac{2\pi}{q}\frac{n_<!}{n_>!}
\left(\frac{q^2}{2}\right)^{n_>-n_<}\big[L_{n_<}^{n_>-n_<}(q^2/2)\big]^2\; ,
\nonumber \\
&& \nonumber \\
&&{\rm otherwise} \;\;\;
v^{N}_{\sigma_1^{\prime},\sigma_2^{\prime},\sigma_1,\sigma_2}=0\; .
\label{vnsamediff}
\end{eqnarray}

If the two pseudospin LLs have opposite real-spins then only scattering
processes that conserve pseudospin (and therefore also real-spin) at each
vertex will contribute to the anisotropy energy.  For $s(\sigma)=
-s(-\sigma)$ and $n(\sigma)=n(-\sigma)\equiv n$ we obtain
\begin{eqnarray}
&&v^N_{\sigma,\sigma,\sigma,\sigma}=v^N_{\sigma,-\sigma,\sigma,-\sigma}=
\frac{2\pi}{q}\big[L_{n}
(q^2/2)\big]^2\; , \nonumber \\
&& \nonumber \\
&&{\rm otherwise} \; \; \;
v^{N}_{\sigma_1^{\prime},\sigma_2^{\prime},\sigma_1,\sigma_2}=0\; .
\label{vndiffsame}
\end{eqnarray}
Finally if the pseudospin LLs
have opposite real-spins {\em and} different orbit-radius quantum
numbers, the in-plane factor in the effective
interactions is
\begin{eqnarray}
&&v^N_{\sigma,\sigma,\sigma,\sigma}=\frac{2\pi}{q}\big[L_{n(\sigma)}
(q^2/2)\big]^2\; , \nonumber \\
&& v^N_{\sigma,-\sigma,\sigma,-\sigma}=\frac{2\pi}{q}L_{n(\sigma)}
(q^2/2)L_{n(-\sigma)}(q^2/2)\; , \nonumber \\
&& \nonumber \\
&&{\rm otherwise} \;\;\;
v^{N}_{\sigma_1^{\prime},\sigma_2^{\prime},\sigma_1,\sigma_2}=0\; .
\label{vndiffdiff}
\end{eqnarray}

Eq.~(\ref{vxi}) for
the subband factor and explicit expressions
(\ref{vnsamesame})-(\ref{vndiffdiff}) for the in-plane factor in
the effective interaction,
together with
Eqs.~(\ref{intq}),(\ref{uij}),(\ref{b}) and Table~\ref{wij}, provide a formal recipe
to calculate the anisotropy coefficients $U_{i,j}$ for crossing LLs with any
combination of quantum well subband, orbit radius and real-spin indeces.
In the following section we use the explicit forms (\ref{vxidiff}) 
and (\ref{vxisame}) for the subband factor to develope a pseudospin 
anisotropy classification scheme for our model bilayer system.
\section{Magnetic anisotropy}

The nature of the anisotropy energy is of qualitative importance
for two-dimensional ferromagnets, including QHFs.
Systems with easy-axis anisotropy, i.e.,
discrete directions at which the energy of the ordered state is
minimized, have long range order at finite temperature and phase
transitions in the Ising universality class.  Systems with
easy-plane anisotropy, i.e., a continuum of coplanar
pseudospin magnetization orientations at which the energy of the
ordered state is minimized, do not have long-range order but do
have Kosterlitz-Thouless phase transitions at a finite
temperature.  In the isotropic case, all directions of pseudospin
magnetization have
identical energy, only the ground state has long-range order, and
there are no finite-temperature phase transitions.  Nevertheless,
magnetization correlations become extremely long at low temperatures.
The objective of this work is to predict which class of QHF
occurs for a particular pair of crossing LLs.

We start our analysis by considering
pseudospin LLs that belong to the
same subband, i.e. $\xi(\uparrow)=\xi(\downarrow)$.
In this case, only LLs with opposite real-spins
can be aligned.
Two examples of QHFs falling into this category,
total filling factor $\nu=1$ with $n(\uparrow)=n(\downarrow)=0$,
and $\nu=2$ with $n(\uparrow)=1$, $n(\downarrow)=0$,
are illustrated in Fig.~\ref{singlelayerll}.  
The cases of $n(\uparrow) \ne n(\downarrow)$  are realized 
when the ratio of the spin-splitting to LL
separation is an integer.  
In GaAs, this 
ratio is only $\sim 1/60$ at perpendicular fields but
can be tuned by tilting
the magnetic field away from the normal to the 2D layer.
For typical well
widths orbital effects of the in-plane field, not included here,
become important  at the tilt angles where the coincidences
of interest are realized and have to be accounted for\cite{mansour}
to obtain  correct values of the pseudospin
anisotropy energy coefficients (\ref{uij}).
However, recent work on AlAs quantum wells\cite{papadakis} 
and InSb quantum wells\cite{santos}
with large Zeeman couplings
have made the situation we study below, which assumes perpendicular
magnetic field, accessible.

When opposite spin LLs cross, Eqs.~(\ref{vndiffsame})
and (\ref{vndiffdiff}) imply that all interactions $B^{\pm}_{i}$ with $i>2$
in (\ref{b}) vanish.  Then the only non-zero anisotropy term is
\begin{eqnarray}
U_{z,z}&=&-\frac18
\int_0^{\infty}dq e^{-q^2/2}\big[L_{n(\uparrow)}(q^2/2)
-L_{n(\downarrow)}(q^2/2)\big]^2
\nonumber \\
&\times&
\big[(1+r_{\Delta}^2)+(1-r_{\Delta}^2)e^{-dq}\big]\; .
\label{uzzonesub}
\end{eqnarray}
Note that the Hartree energy contribution to anisotropy always vanishes when
the crossing LLs share the same subband wavefunction.
If the two pseudospin levels also have the same orbit-radius quantum number then
Eq.~({\ref{uzzonesub}) gives
$U_{z,z}=0$ and the ferromagnetic state is {\em isotropic}.
Physically, the result follows from the independence
of the Coulomb interaction strength on real-spin.
An important example of these isotropic QHFs occurs when $n(\uparrow)=
n(\downarrow)=0$ and $r_{\Delta}=1$, i.e., there is no tunneling between
layers. This is the thoroughly studied single-layer $\nu=1$
QHF\cite{dassarmabook,qhferro,skyrmexps} for which the HF theory
ground state happens to be exact.  We remark that 
quantitative estimates based on the HF mean-field theory
presented here 
require corrections to quantum fluctuation effects in cases when
the ordered pseudospin moment direction is not a good
quantum number. 
A detail understanding of these corrections is one challenge for 
future experimental and theoretical work on QHFs.

For $n(\uparrow)\neq n(\downarrow)$,
Eq.~(\ref{uzzonesub}) implies that $U_{z,z}<0$,
making the $z$-axis the easy pseudospin orientation axis.
Again, at $r_{\Delta}=1$ our model reduces to that of a 
single layer 2D systems whose {\em easy-axis} anisotropy at even
filling factors has been identified previously.\cite{mansour}
In finite-thickness single quantum wells, a QHF with pseudospin LLs of
the same subband but different real-spin and orbit-radius indices is also
easy-axis.  The magnitude of the anisotropy will decrease with layer thickness,
as can be seen by comparing $U_{z,z}$ in (\ref{uzzonesub}) calculated for
$r_{\Delta}=1$ and $r_{\Delta}=0$. (Note that the single-subband
unbiased double well with finite tunneling, i.e. $r_{\Delta}=0$, models a single layer system
with an effective thickness $d$.)

We now turn to the crossing of LLs with different subband 
indices, for which the pseudospin anisotropy physics is richer.
In Fig.~\ref{doublelayerll1} we show examples of $n(\uparrow)=n(\downarrow)$
bilayer QHFs for $\nu=1$
and $\nu=2$ based on same real-spin and opposite
real-spin LLs respectively.
Eqs.~(\ref{uij}),(\ref{intq}),(\ref{vxidiff}),(\ref{vnsamesame}),
and Table~\ref{wij} imply four non-zero anisotropy terms for
$n(\uparrow)=n(\downarrow)\equiv n$ and $s(\uparrow)=s(\downarrow)$:
\begin{eqnarray}
&&U_{z,z}=ur_{\Delta}^2\; , \nonumber \\
&&U_{x,x}=u(1-r_{\Delta}^2)\; , \nonumber \\
&&U_{x,z}=U_{z,x}=ur_{\Delta}(1-r_{\Delta}^2)^{1/2}\; ,
\nonumber \\
&&
\nonumber \\
&& {\rm where}
\nonumber \\
&&
\nonumber \\
&&u=\frac d2-\frac12
\int_0^{\infty}dq e^{-q^2/2}\big[L_n(q^2/2)\big]^2
\big(1-e^{-dq}\big)\; .
\label{utwosub}
\end{eqnarray}
First term in the expression for energy $u$ comes from the Hartree interaction,
the second term represents exchange contribution which is always smaller
than the Hartree energy in this case, i.e., $u>0$. For zero tunneling between
layers ($r_{\Delta}=1$), the only non-zero anisotropy energy component,
$U_{zz}=u$, is positive
leading to the {\em easy-plane} anisotropy of the QHF.
In the absence of symmetry breaking fields (the linear pseudospin magnetization
terms in (\ref{ehf}))
the variational energy (\ref{ehf})
is minimized when the pseudospins condense into a state magnetized at 
an arbitrary orientation within 
in the $x-y$ plane. In this state, electronic charge is distributed equally between
the layers (pseudospin angle $\theta=0$)
minimizing the electrostatic energy; spontaneous interlayer phase coherence\cite{dltheory}
the physical counter part of pseudospin order in this case, 
lowers the total energy of the system by strengthening interlayer 
exchange interactions.

For $r_{\Delta}^2<1$, Eqs.~(\ref{utwosub}) imply the following quadratic
terms in the HF total energy
\begin{equation}
\sum_{i,j=x,y,z} U_{i,j}m_im_j=u\big[r_{\Delta}
m_z+(1-r_{\Delta}^2)^{1/2}m_x\big]^2\; ,
\label{utwosubrd}
\end{equation}
i.e., the easy-plane is tilted from the $x-y$ plane in the
pseudospin space by angle $\alpha=\arctan
\big[(1-r_{\Delta}^2)^{1/2}/r_{\Delta}\big]$. The pseudospin basis 
states at different values of $r_{\Delta}$ are related, however, by a unitary
transformation, which corresponds precisely to a rotation about $y$-axis by angle $-\alpha$, as
seen from Eq.~(\ref{zwf}).  The easy-plane where the above anisotropy
energy is constant is, for any value of $r_{\Delta}$, the plane of equal charge per layer.

When pseudospin-up and pseudospin-down states
differ by more than their subband indices, by their spin indices for example,
the two pseudospin basis sates at different $r_{\Delta}$ are not related by
a unitary transformation.  In this case the magnetic anisotropy does depend
on $r_{\Delta}$.   For example, consider the case  
$\xi(\uparrow)=-\xi(\downarrow)$,
$n(\uparrow)=n(\downarrow)$, and $s(\uparrow)=-s(\downarrow)$.
For opposite real-spin LLs, all anisotropy terms that include pseudospin
non-conserving scattering processes
drop out. The only non-zero energy term, $U_{z,z}$, has the same value as
in~(\ref{utwosub}). Hence, the QHF is {\em isotropic}
in the unbiased ($r_{\Delta}=0$) bilayer system
while applying external bias ($r_{\Delta}>0$) leads to {\em easy-plane}
anisotropy in pseudospin space.

At this point let us make an experimentally important comment on 
the bilayer systems realized
in wide single quantum wells.\cite{dassarmabook,kt,wsqw}
The difference between this sample
geometry and the double quantum well with narrow (in our
model infinitely narrow) layers is in the nature of the barrier responsible
for the bilayer character of the
electronic system. In wide quantum wells the barrier
is soft,\cite{wsqw} originating
from Coulomb interactions among electrons in the well.
Then the tunneling probability
between layers is strongly dependent on the electron density and  
quantum well subband populations. This tunability makes wide single
quantum wells an experimentally
attractive alternative to double wells in studies of bilayer quantum
Hall phenomena. The softness of the barrier can, however, lead to qualitatively
important consequences for the ordered many-particle states. Translated
into the pseudospin language, $\Delta_t$ cannot be treated as an
external one-body field
acting on the pseudospin particles but, in general, will
depend\cite{vittorio} on the
pseudospin orientation in the ordered ground state. For the 
pseudospin LLs discussed in the previous paragraph
($\xi(\uparrow)=-\xi(\downarrow)$,
$n(\uparrow)=n(\downarrow)$, and
$s(\uparrow)=-s(\downarrow)$) and for $r_{\Delta}=0$,
this effect can lead to an anisotropic QHF.
LSDA calculations indicate that at low electron
densities the anisotropy will be easy-plane while easy-axis anisotropy
is more likely to
develop at high densities.\cite{vittorio}
We make this remark to point out that not
all results obtained for the double quantum well model are  directly
applicable to bilayers in wide single wells. In many cases the
theoretical description
of QHFs  in  wide single wells
requires modifications of the idealized bilayer model to account for
mixing of higher electrical subbands.
The self-consistent LSDA  for the growth
direction single-particle orbitals is a particularly convenient, if 
somewhat {\it ad hoc}, method that allows any sample geometry to be studied
while retaining the basic structure of the many-body HF formalism for QHFs.

In the remaining part of this section we consider pseudospin LLs with
opposite subband indices {\em and} different orbit radius quantum numbers.
At total filling factor $\nu=3$, for example,
the pseudospin LLs will have the same real-spin while at $\nu=4$ opposite
spin LLs can be aligned, as shown in Fig.~\ref{doublelayerll2}.
A common feature of the QHFs discussed below is the transition from a 
state with easy-axis anisotropy to a state with easy-plane anisotropy as
$r_{\Delta}$ and the layer separation $d$ are varied.
For $s(\uparrow)=s(\downarrow)$,
Eqs.~(\ref{uij}),(\ref{intq}),(\ref{vxidiff}),(\ref{vnsamediff}),
and Table~\ref{wij} give three non-zero anisotropy energies
\begin{eqnarray}
&&U_{z,z}=\frac{r_{\Delta}^2d}{2}-
\frac18
\int_0^{\infty}dq e^{-q^2/2}\nonumber \\
&&\times\bigg\{\big[L_{n(\uparrow)}(q^2/2)
-L_{n(\downarrow)}(q^2/2)\big]^2
\big(1+e^{-dq}\big)\nonumber \\
&&+
r_{\Delta}^2\big[L_{n(\uparrow)}(q^2/2)+L_{n(\downarrow)}(q^2/2)\big]^2
\big(1-e^{-dq}\big)\bigg\}
\; , \nonumber \\
&&U_{x,x}=U_{y,y}=-\frac14\int_0^{\infty}dq e^{-q^2/2}
\frac{n_<!}{n_>!}
\left(\frac{q^2}{2}\right)^{n_>-n_<}\nonumber \\
&&\times\big[L_{n_<}^{n_>-n_<}(q^2/2)\big]^2
\big(1-r_{\Delta}^2\big)\big(1-e^{-dq}\big)\; ,
\nonumber \\
&&
\nonumber \\
&& {\rm where}
\nonumber \\
&&
\nonumber \\
&& n_{<}\equiv\min\big[n(\uparrow),n(\downarrow)\big]\;,\;\;
n_{>}\equiv\max\big[n(\uparrow),n(\downarrow)\big]
\; .
\label{utwosub2}
\end{eqnarray}
The HF total energy contributions that are
quadratic in the pseudospin magnetization
components can be grouped as
\begin{equation}
\sum_{i,j=x,y,z} U_{i,j}m_im_j=(U_{z,z}-U_{x,x})m_z^2+U_{x,x}\;.
\label{utwosubrd2}
\end{equation}
(Recall that $\hat m$ is a unit vector, i.e. $m_x^2+m_y^2=1-m_z^2$.)
From Eq.~(\ref{utwosubrd2}) we obtain that
for $U_{z,z}-U_{x,x}<0$ the QHF has {\em easy-axis} anisotropy while for
$U_{z,z}-U_{x,x}>0$ the system is an {\em easy-plane} ferromagnet. At the
critical layer separation $d=d^*$, obtained from the condition
$U_{z,z}=U_{x,x}$, the magnetic
anisotropy vanishes and a fine-tuned isotropy is achieved.
It follows from Eqs.~(\ref{utwosub2}) that $d^*$ is finite
for all values of $r_{\Delta}$.

For pseudospin LLs with
$s(\uparrow)=-s(\downarrow)$, the anisotropy energy components $U_{x,x}$ and
$U_{y,y}$ vanish and the critical layer separation $d^*$ corresponds to
$U_{z,z}=0$, where $U_{z,z}$ is given by the same expression as in the
$s(\uparrow)=s(\downarrow)$ case (see Eq.~(\ref{utwosub2})).
Since $U_{z,z}<0$ at $r_{\Delta}=0$,
the critical separation $d^*$ diverges in the absence of external
bias, i.e., easy-axis pseudospin anisotropy does not exist for any layer separation.
For $r_{\Delta}>0$ the transition between
easy-plane and easy-axis anisotropy occurs at finite $d$ as for the
$s(\uparrow)=s(\downarrow)$ pseudospin LLs.

In Figs.~\ref{phasediag}(a) and \ref{phasediag}(b) we show the magnetic anisotropy
phase diagrams in the $d$-$r_{\Delta}$ plane calculated for  $\nu=3$
and $\nu=4$ QHFs (see Fig.~\ref{doublelayerll2}). Since the layer separation is
in units of magnetic length, these figures imply that 
transition between easy-axis and easy-plane anisotropies at a given filling factor
can be induced in one physical sample by changing the density
of the 2D electron system. High electron densities would
correspond to the easy-plane region, and low densities to the easy-axis region.
Note that these numerical results confirm the general remark made above, since the
critical layer diverges as $r_{\Delta} \to 0$ for $\nu=4$ while it 
remains finite for $\nu=3$.

\section{Symmetry breaking fields}
The pseudospin orientation in a QHF ground state is determined by minimizing
the variational total energy~(\ref{ehf}). In the absence of energy
terms that are linear in pseudospin magnetization components, the
HF ordered states spontaneously break continuous SU(2) or U(1)
symmetry\cite{dltheory} in the case of
isotropic or easy-plane QHFs respectively,
and the discrete symmetry between pseudospin-up and
pseudospin-down orientations in the case of easy-axis QHFs.
In this section we take into account external and internal potentials
which contribute to the linear terms in the HF total energy
and comment on the pseudospin reversal that can be triggered by
adjusting these symmetry breaking fields.
We focus on an case which we feel is particularly appropriate for 
experimental study by considering the ordered $\nu=3$ quantum Hall state.
Similar considerations would apply for all classes of QHFs discussed
in this paper.

The pseudospin LLs in the bilayer $\nu=3$ QHF (see
Fig.~\ref{doublelayerll2}) have opposite subband indices and 
orbit radius quantum numbers
$n=0$ and $n=1$, respectively. We call the [$\xi=-1$,$n=0$,$s=+1/2$]
LL the pseudospin-up state
and the [$\xi=1$,$n=1$,$s=+1/2$] LL the pseudospin-down state. With this definition,
the one-body potentials in~(\ref{ehf}) can be written as
\begin{eqnarray}
b_z&=&-\frac12\big[(\Delta_V^2+\Delta_t^2)^{1/2}-\hbar\omega_c+I_F
-I_{H,z}\big]\; ,\nonumber \\
b_x&=&\frac12I_{H,x}\; , \nonumber \\
b_y&=&0\; .
\label{bxyz}
\end{eqnarray}
The effective field $I_F$ is the difference between pseudospin-up
and pseudospin-down particle exchange energy with electrons in the fully
occupied [$\xi=1$,$n=0$,$s=+1/2$] LL, i.e,
\begin{eqnarray}
I_F&=&\frac12\int_0^{\infty}dqe^{-q^2/2}\bigg\{\frac{q^2}{2}\big[1+
r_{\Delta}^2+(1-r_{\Delta}^2)e^{-dq}\big]\nonumber\\
&-&(1-r_{\Delta}^2)
(1-e^{-dq})\bigg\}\; .
\label{if}
\end{eqnarray}
For $r_{\Delta}>0$,
electrons in the [$\xi=1$,$n=0$,$s=\pm1/2$] LLs produce also an electrostatic
field, represented by  $I_{H,z}$ and $I_{H,x}$ in~(\ref{bxyz}),
which screens the external bias potential. Since this effective
field favors occupation of a particular layer rather than a particular
pseudospin state it  couples to $z$ and $x$ components of the pseudospin
operator. The energy inbalance between the two layers produced by $\vec{
I}_H$ is $2dr_{\Delta}$ which, together with Eq.~(\ref{zwf}), gives
\begin{eqnarray}
I_{H,z}&=&2dr_{\Delta}^2\; ,\nonumber \\
I_{H,x}&=&2dr_{\Delta}(1-r_{\Delta}^2)^{1/2}\; .
\label{ihzx}
\end{eqnarray}
In the expression for the HF total energy~(\ref{ehf}), we 
included explicitly the contribution to the symmetry breaking fields
which results from  Coulomb interactions between electrons
in the pseudospin LLs.
For the $\nu=3$ QHF we are considering here, only $U_{{\protect\bf 1},z}$ and
$U_{z,{\protect\bf 1}}$ energies are non-zero:
\begin{eqnarray}
U_{{\protect\bf 1},z}=U_{z,{\protect\bf 1}}&=&
-\frac18\int_0^{\infty}dqe^{-q^2/2}\big[1+
r_{\Delta}^2+(1-r_{\Delta}^2)e^{-dq}\big]\nonumber\\
&\times&(q^2-q^4/4)\; .
\label{u1z}
\end{eqnarray}

In the bilayer system with no tunneling ($r_{\Delta}=1$), $I_{H,x}=0$ and
the total symmetry breaking field,
$b^*\equiv b_z-U_{{\protect\bf 1},z}$, is oriented along the $z$ pseudospin
direction and is given by
\begin{eqnarray}
b^*&=&
-\frac12\big[(\Delta_V^2+\Delta_t^2)^{1/2}-\hbar\omega_c+\sqrt{\pi/2}/2
-2d\nonumber \\
&-&\sqrt{\pi/2}/8\big]\;.
\label{bstar}
\end{eqnarray}
In Figs.~\ref{psreversal}(a)-(c)
we plot the pseudospin evolution with effective field $b^*$ for the three anisotropy
regimes of a $\nu=3$ QHF with $r_{\Delta}=1$. At the phase boundary
between easy-axis and easy-plane anisotropies, the $\nu=3$ QHF is isotropic
and the pseudospin reverses abruptly at $b^*=0$ (see Fig.~\ref{psreversal}(a)).
In the easy-plane anisotropy regime, ($U_{z,z}-U_{x,x}>0$), the pseudospin
evolves continuously with $b^*$ as illustrated in Fig.~\ref{psreversal}(b)
reaching alignment with $b^*$ at $|b^*|\ge U_{z,z}-U_{x,x}$.
For the easy-axis anisotropy case ($U_{z,z}-U_{x,x}<0$), the HF energy has two local
minima at  $m_z=\pm 1$ when $|b^*|<|U_{z,z}-U_{x,x}|$. The pseudospin-up and
pseudospin-down polarized states are separated by an energy barrier which
results in the hysteretic pseudospin-reversal behavior shown
in Fig.~\ref{psreversal}(c).

In bilayer systems with non-zero tunneling, pseudospin reversal
follows a more complicated pattern in which the competition
between $x$ and $z$ components of the symmetry breaking field plays
an important role.  In general, the pseudospin will rotate in
the $x$-$z$ plane, i.e. $m_x=\sqrt{1-m_z^2}$.  Since
the derivative of the HF energy with respect to $m_z$ diverges at $m_z=\pm1$
due to the $I_{H,x}$ term the pseudospin will never align
completely with the $z$-axis when $r_{\Delta}<1$.
\section{Summary}

In the strong magnetic field limit, the physics of high mobility 
two-dimensional electron systems is usually dominated by 
electron-electron interactions except at integer filling factors, where the 
single-particle physics responsible for the gap between Landau levels assumes
the dominant role.  When external parameters are adjusted so that 
two or more Landau levels simultaneously approach the chemical potential,
the integer
filling factor case is less exceptional, interaction effects are 
always strong, and uniform density broken symmetry ground states analogous
to those in conventional ferromagnets are common.  In this paper we have 
discussed how the nature of these states depends on the character of the 
nearly degenerate Landau levels.   Our attention is restricted to the case
where only two Landau levels are close to the chemical potential and 
we distinquish these crossing Landau levels by introducing a pseudospin
degree of freedom.  

Using Hartree-Fock variational wavefunctions, we are able to derive
an explicit expression for the dependence of ground state energy on pseudospin
orientation for crossing LLs with any combination
of quantum well subband, orbit radius and real-spin
degree-of-freedom quantum numbers.  As in conventional magnetic systems,
qualitative differences exist between the physical properties of isotropic
(Heisenberg) systems with no dependence of energy on pseudospin orientation,
easy-axis (Ising) systems with discrete prefereed pseudospin orientations,
and (XY) easy-plane systems for which the minimum is achieved simultaneously
for a plane of orientations.  Our mean-field  results predict which class of 
pseudospin quantum Hall ferromagnet occurs in different circumstances.
We focus on a model commonly used for bilayer
quantum Hall systems in which the finite width of both quantum wells 
is neglected.  The external parameters of the model are the 
Zeeman coupling strength, the bias potential between the wells 
$\Delta_V$, and the single particle splitting due to interlayer tunneling $\Delta_t$.
In the limit $\Delta_t=0$, this model applies to a single quantum
well when the crossing Landau levels have the same 
subband wavefunction, i.e., are in the same quantum well. 
Classification predictions for this model as a function of 
layer separation $d$ and the ratio $r_{\Delta} = \Delta_V/
(\Delta_V^2+\Delta_t^2)^{1/2}$  
are summarized in the diagram~\ref{class}. 

For the single quantum well case we find 
isotropic behavior when the orbit radius 
quantum numbers of the crossing Landau levels are identical, and 
easy-axis behavior otherwise.  In general when the crossing 
Landau levels have identical subband wavefunctions, the nature of the 
pseudospin anisotropy  does not depend on the paramters 
$d$ and $r_{\Delta}$.
For different subband wavefunctions the pseudospin anisotropy can
vary in the $d-r_{\Delta}$ plane.
Particularly intriqueing is the case of crossing LLs with different subband
and orbit radius quantum
numbers where, at a given filling factor, the system  can undergo
a quantum phase transition from an easy-axis to easy-plane QHF. 
At the
phase boundary the pseudospin anisotropy vanishes and a fine-tuned
isotropy is achieved.  The critical values of $d$ and $r_{\Delta}$ are
experimentally accesible and may be accompanied by 
observable changes in pseudospin reversal properties as external
parameters are varied.

\section*{Acknowledgments}

This work was supported by the National Science Foundation under
grants DMR-9623511, DMR-9714055, and DGE-9902579 and
by the Grant Agency of the Czech Republic
under grant 202/98/0085.

\begin{table}
\begin{center}
\begin{tabular}{ccccc}
 & ${\protect\bf 1}$
& $x$ & $y$ & $z$
\\
${\protect\bf 1}$ & $B_1^++B_2^+$ &  $B_5^-+B_6^+$ &
$\frac1iB_5^-+\frac1iB_6^-$& $B_1^--B_2^-$\\
$x$ & $B_7^++B_8^+$ & $B_3^++B_4^+$ &
$\frac1iB_3^--\frac1iB_4^-$ & $B_7^+-B_8^+$ \\
$y$ & $\frac1iB_7^-+\frac1iB_8^-$ & $\frac1iB_3^-+\frac1iB_4^-$ &
$B_3^+-B_4^+$&$\frac1iB_7^--\frac1iB_8^-$  \\
$z$ &$B_1^-+B_2^-$  &$B_5^+-B_6^+$  &
$\frac1iB_5^--\frac1iB_6^-$&$B_1^+-B_2^+$
\end{tabular}
\end{center}
\caption{\protect Coulomb interaction matrix elements $W_{i,j}$; $i,j=
{\protect\bf 1},x,y,z$. The $B^{\pm}_n$ terms are defined in (\ref{b}).
}
\label{wij}
\end{table}

\begin{figure}
\caption{Schematic of the conduction band edge profile of a biased
double-quantum-well sample with non-zero tunneling between 2D layers. Energy
levels due to quantization of electron motion along the growth direction are
also indicated.  For concrete calculations we assume infinitely narrow quantum wells
separated by the distance $d$.
}
\label{dqw}
\end{figure}

\begin{figure}
\caption{Schematic LL diagrams for $\nu=1$ (a) and $\nu=2$ (b)
single-subband
QHFs. Index $\sigma=\uparrow,\downarrow$ labels the pseudospin LLs.
The crossing LLs are indicated by half-solid dots, while
inert filled LLs are indicated by solid dots and inert empty LLs by open dots.
}
\label{singlelayerll}
\end{figure}

\begin{figure}
\caption{Schematic LL diagrams for $\nu=1$ (a) and $\nu=2$ (b)
bilayer
QHFs. The pseudospin LLs have opposite subband indices and same (a)
or opposite (b) real-spins.
}
\label{doublelayerll1}
\end{figure}

\begin{figure}
\caption{Schematic LL diagrams for $\nu=3$ (a) and $\nu=4$ (b)
bilayer
QHFs. The pseudospin LLs have opposite subband indices,
different orbit radius quantum numbers and  same (a)
or opposite (b) real-spins.
}
\label{doublelayerll2}
\end{figure}

\begin{figure}
\caption{Magnetic anisotropy phase diagrams
for bilayer $\nu=3$ (a) and $\nu=4$ (b) QHFs
from Fig.~\protect\ref{doublelayerll2}.
In the white region the anisotropy is
easy-axis, in the grey region the QHF has easy-plane anisotropy.
At the phase boundary
the ferromagnetic state is isotropic.
}
\label{phasediag}
\end{figure}

\begin{figure}[h]
\caption{Pseudospin orientation as a function of the effective
field $b^*$ for the isotropic QHF and as a function of $b^*$ relative
to the anisotropy energy $|U_{z,z}-U_{x,x}|$  for the easy-plane (b),
and easy-axis (c) QHFs at $\nu=3$.
}
\label{psreversal}
\end{figure}

\begin{figure}
\caption{Magnetic anisotropy of QHFs with pseudospin LL subband
indices $\xi(\sigma)$, orbit radius quantum numbers $n(\sigma)$, and
real-spins $s(\sigma)$.
}
\label{class}
\end{figure}

\begin{references}
\bibitem{ahmintro}
A.H. MacDonald,  in {\em Proceedings of the 1994 Les Houches Summer
School on Mesoscopic Quantum Physics}, edited by E. Akkermans
{\it et~al.} (Elsevier Science, Amsterdam, 1995), pp.\ 659--720.

\bibitem{dassarmabook} For a review on QHFs at $\nu=1$ see
experimental chapter by J.P. Eisenstein and theoretical chapter by
S.M. Girvin and A.H. MacDonald in
{\it Perspectives on Quantum Hall Effects} (Wiley, New York, 1997).

\bibitem{qhferro} C. Kallin and B.I. Halperin, Phys. Rev. B {\bf 30},
5655 (1984);
D.H. Lee and C.L. Kane, Phys. Rev. Lett. {\bf 64},
1313 (1990);
S.L. Sondhi, A. Karlhede, S.A. Kivelson, and E.H.
Rezayi, Phys. Rev. B {\bf 47}, 16419 (1993);
A.H. MacDonald, H.A. Fertig, and L. Brey, Phys. Rev. Lett.
{\bf 76}, 2153 (1996).
\bibitem{skyrmexps} S.E. Barrett, G. Dabbagh, L.N. Pfeiffer, and K.W. West,
Phys. Rev. Lett. {\bf 74}, 5112, (1995); R. Tycko, S.E. Barrett, G. Dabbagh,
L.N. Pfeiffer, and K.W. West, Science {\bf 268}, 1460 (1995);
A. Smeller, J.P. Eisenstein,  L.N. Pfeiffer, and K.W. West,
Phys. Rev. Lett. {\bf 75}, 4290 (1995);
D.K. Maude, M. Potemski, J.C. Portal, M. Henini, L. Eaves, G. Hill,
and M.A. Pate, Phys. Rev. Lett. {\bf 77}, 4604 (1996);
E.H. Aifer, B.B. Goldberg, and D.A. Broido, Phys. Rev. Lett. {\bf 76},
680 (1996);
V. Bayot, E. Grivei, J.M. Beuken, S. Melinte, and M. Shayegan,
Phys. Rev. Lett. {\bf 76}, 4584 (1996); {\bf 79}, 1718 (1997).

\bibitem{ahm90} A.H. MacDonald, P.M. Platzman,
G.S. Boebinger, Phys. Rev. Lett. {\bf 65}, 775 (1990);
\bibitem{dltheory}
T. Chakraborty and Pietil\"ainen, {\em ibid} {\bf 59}, 2784 (1987);
X.G. Wen and A. Zee,
Phys. Rev. B {\bf 47}, 2265 (1993); Z.F. Ezawa and A. Iwazaki,
Int. J. Mod. Phys. B {\bf 19}, 3205 (1992); L. Brey, Phys. Rev. Lett. {\bf 65},
903 (1990); H.A. Fertig, Phys. Rev. B {\bf 40}, 1087 (1989); O. Narikiyo
and D. Yoshioka, J. Phys. Soc. Jap. {\bf 62}, 1612 (1993); R.
C$\hat{\rm o}$t\'e, L. Brey, and A.H. MacDonald, Phys. Rev. B {\bf 46},
10239 (1992); X.M. Chen and J.J. Quinn, {\em ibid.} {\bf 45}, 11 054 (1992);
K. Moon {\em et al.}, Phys. Rev. B {\bf 51}, 5138 (1995);
K. Yang {\em et al.}, {\em ibid.} {\bf 54}, 11 644 (1996).

\bibitem{eisensteinlatest} I.B. Spielman, J.P. Eisenstein, L.N. Pfeiffer,
and K.W. West, cond-mat/0002387.

\bibitem{murphy} S.Q. Murphy, J.P. Eisenstein, G.S. Boebinger,
L.N. Pfeiffer, and K.W. West, Phys. Rev. Lett. {\bf 72}.
728 (1994);

\bibitem{kt} T.S. Lay, Y.W. Suen, H.C. Manoharan, X. Ying,
M. Santos, and M. Shayegan, {\em ibid.} {\bf 50}, 17725 (1994);
M. Abolfath, R. Golestanian, and T. Jungwirth,  Phys. Rev. B {\bf 61},
4762 (2000).

\bibitem{mansour} T. Jungwirth, S.P. Shukla, L. Smr\v{c}ka,
M. Shayegan, and A.H. MacDonald, Phys. Rev. Lett. {\bf 81}, 2328 (1998).

\bibitem{vittorio} V. Piazza, V. Pellegrini, F. Beltram, W. Wescheider,
T. Jungwirth, and A.H. MacDonald, Nature {\bf 402}, 638 (1999).

\bibitem{giuliani} G.F. Giuliani and J.J. Quinn, Phys. Rev. B{\bf 31},
6228 (1985).

\bibitem{woowon} Recently observed hysteresis at $\nu=2/5$ and 4/9,
corresponding to composite fermion filling factors $\nu=2$ and 4, has
also been
interpreted within the easy-axis QHF framework: H. Cho, J.B. Young,
W. Kang, K.L. Campman, A.C. Gossard, M. Bichler, and W. Wescheider,
 Phys. Rev. Lett. {\bf 81}, 2522 (1998).

\bibitem{fourcomptheory} L. Zheng, R.J. Radtke, and S. Das Sarma,
Phys. Rev. Lett. {\bf 78}, 2453 (1997); S. Das Sarma, S. Sachdev and L. Zheng,
Phys. Rev. Lett. {\bf 79}, 917 (1997); Phys. Rev. B {\bf 58}, 4672 (1998);
L. Brey, E. Demler, S. Das Sarma, Phys. Rev. Lett. {\bf 83}, 168 (1999);
B. Paredes, C. Tejedor, L. Brey, and L. Mart\'{\i}n-Moreno,
Phys. Rev. Lett. {\bf 83}, 2250 (1999).

\bibitem{fourcompexp} V. Pellegrini, A. Pinczuk, B.S. Dennis, A.S. Plaut,
L.N. Pfeiffer, and K.W. West, Phys. Rev. Lett. {\bf 78}, 310 (1997);
Science {\bf 281}, 799 (1998);
A. Sawada, A.F. Ezawa, H. Ohno, Y. Horikoshi,
Y. Ohno, S. Kishimoto, F. Matsukura, Phys. Rev. Lett. {\bf 80}, 4534 (1998).

\bibitem{bastard} G. Bastard, {\em Wave Mechanics Applied to Semiconductor
Heterostructures} (Les \'{E}ditions de Physique, Paris, 1990).

\bibitem{papadakis} S.J. Papadakis, E.P. De Poortere, and M. Shayegan,
Phys. Rev. B {\bf 59}, R12743 (1999).

\bibitem{santos} K.J. Goldammer, S.J. Chung, W.K. Liu, M.B. Santos,
J.L. Hicks, S. Raymond, and S.Q.Murphy, J. Cryst. Growth {\bf 202}, 753 (1999).

\bibitem{wsqw} Y.W. Suen, J. Jo, M.B. Santos, L.W. Engel, S.W. Hwang, and
M. Shayegan, Phys. Rev. B {\bf 44}, 5947 (1991);
T.S. Lay, T. Jungwirth, L. Smr\v{c}ka, M. Shayegan, Phys. Rev. B {\bf 56},
R7092 (1997).
\end{references}
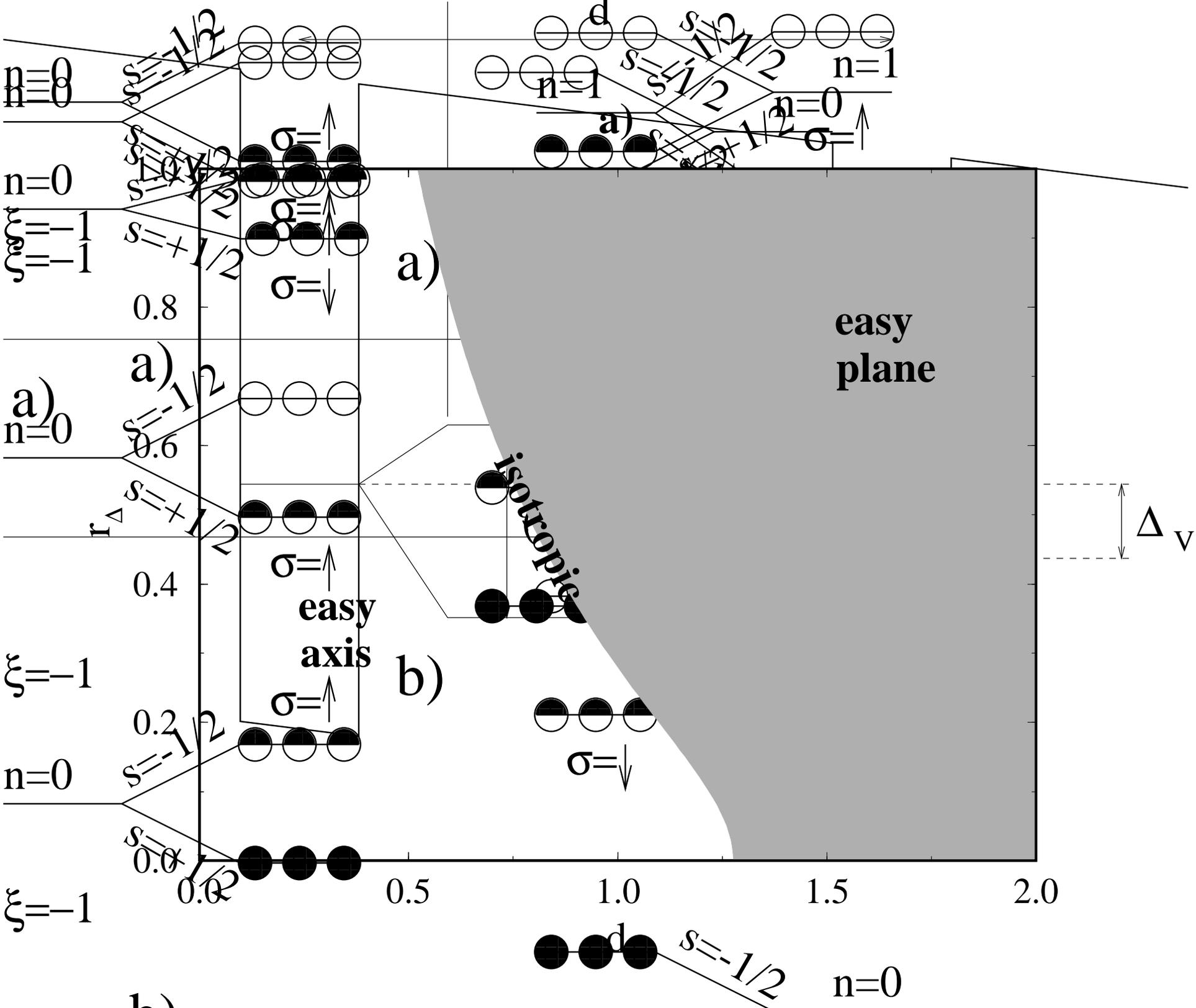
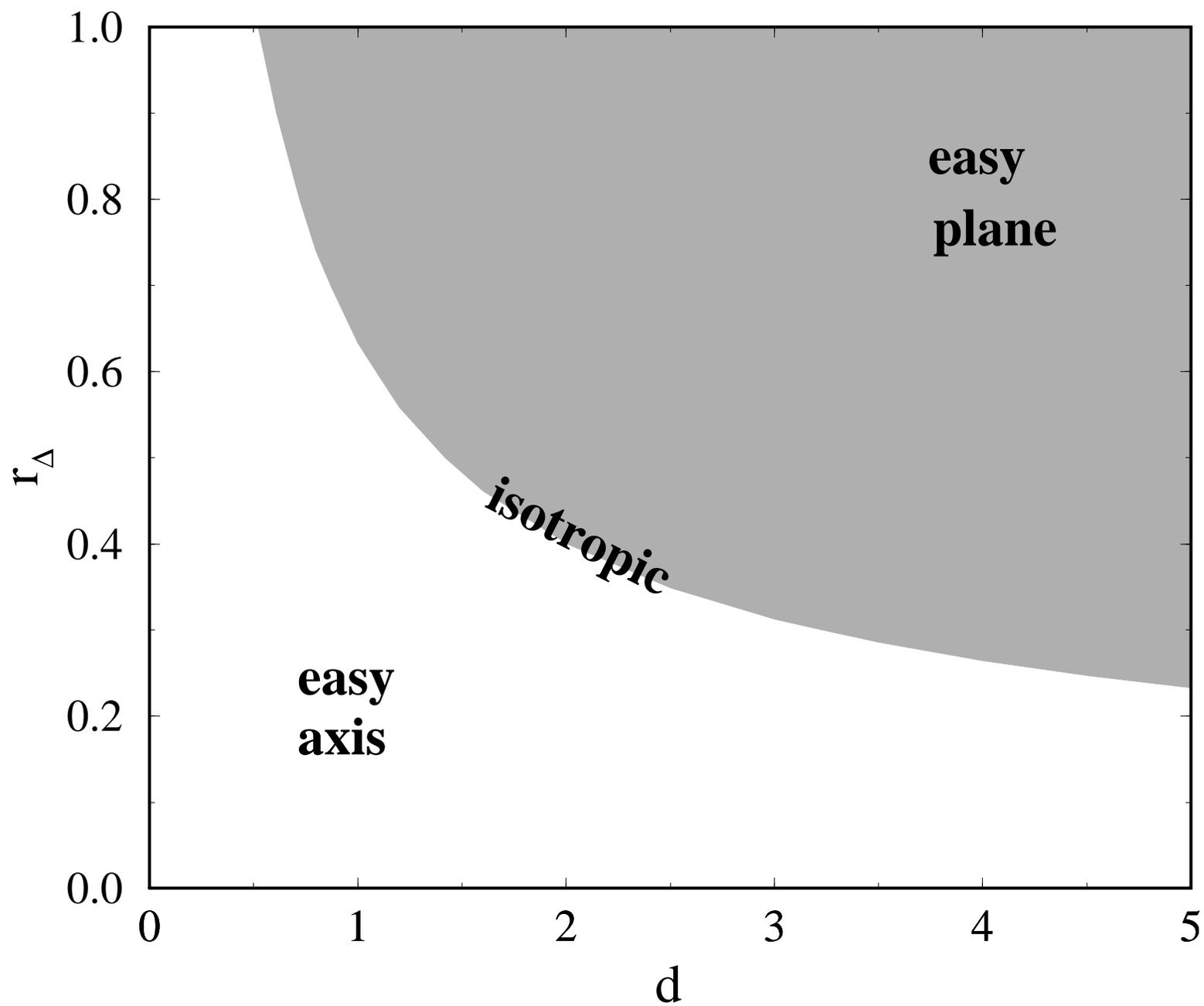
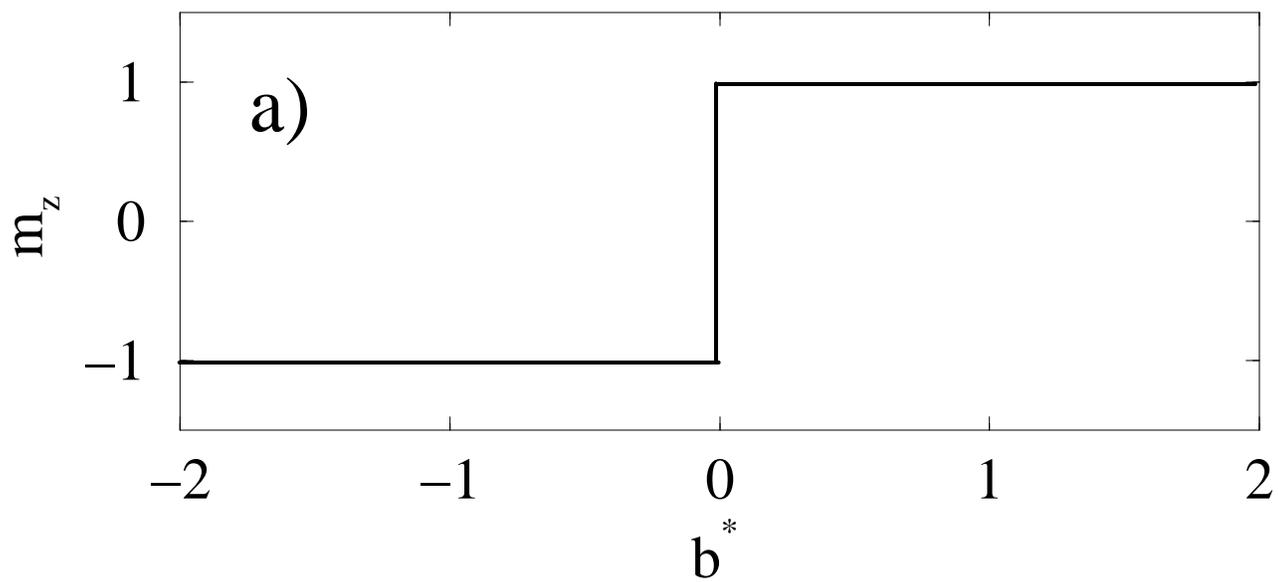
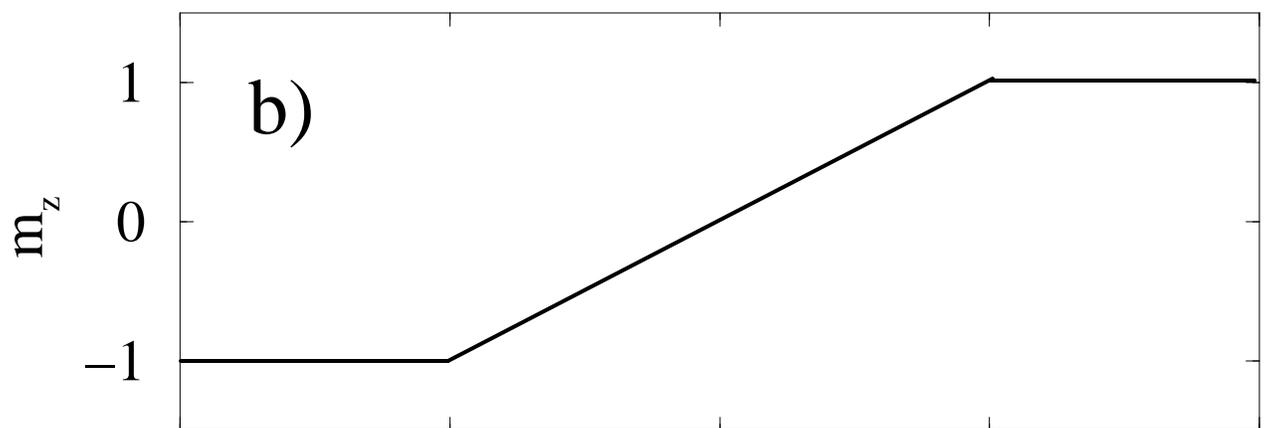
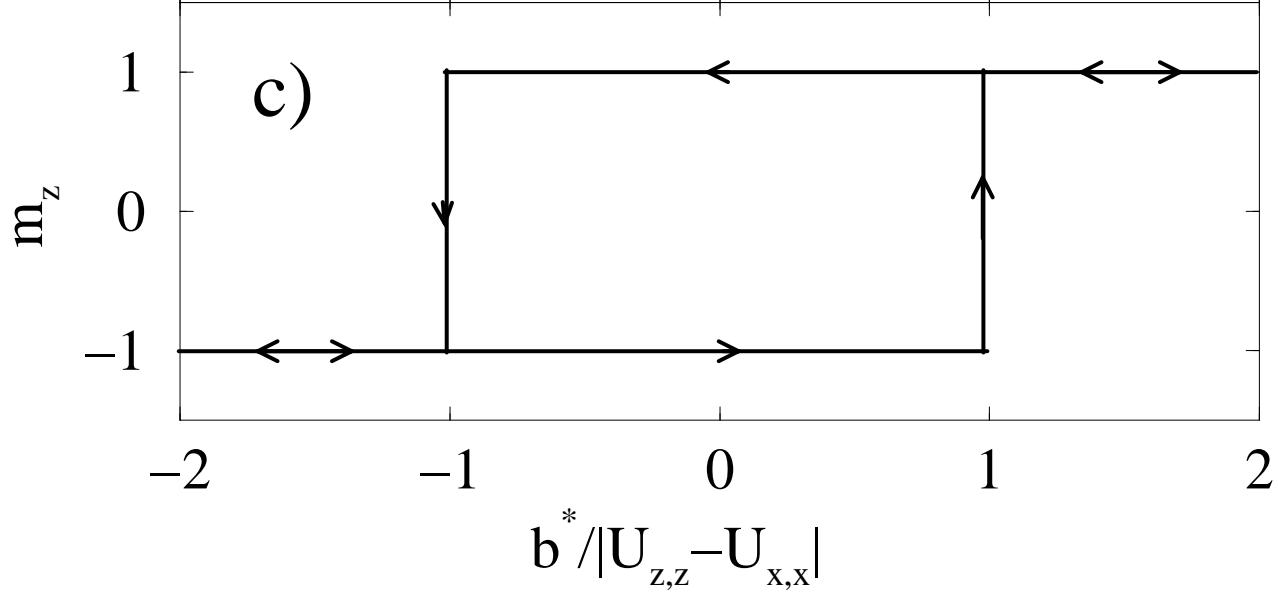
\end{document}